# Probing the multi spin-phonon coupling and local B-site disorder in $Pr_2CoFeO_6$ by Raman spectroscopy and correlation with its electronic structure by X-ray photoemission spectroscopy

Arkadeb Pal[1], Surajit Ghosh[1], Amish G. Joshi[2], P. K. Gupta[1], P. Prakash[3], Amitabh Das[3,4], A. K. Ghosh[5], and Sandip Chatterjee[1#]

[1]Indian Institute of Technology (BHU) Varanasi 221005, India

[2]CSIR-National Physical Laboratory, Dr. K. S. Krishnan Road, New Delhi 110012, India

[3]Solid State Physics Division, Bhaba Atomic Research Centre, Mumbai 400085, India

[4]Homi Bhabha National Institute, Anushaktinagar, Mumbai 400094, India

[5]Banaras Hindu University, Varanasi 221005, India

## Abstract

Electronic structure near Fermi level of $Pr_2CoFeO_6$ (at 300 K) was investigated by X-ray photoemission spectroscopy (XPS) technique. All three cations, i.e., Pr, Co and Fe were found to be trivalent in nature. XPS analysis also suggested the system to be insulating in nature. Moreover, Raman spectroscopy study indicated the random distribution of the B-site ions (Co/Fe) triggered by same charge states. In temperature-dependent Raman study, the relative heights of the two observed phonon modes exhibited anomalous behaviour near magnetic transition temperature $T_N$~270 K, thus indicating towards interplay between spin and phonon in the system. Furthermore, clear anomalous softening was observed below $T_N$ which confirmed the existence of strong spin-phonon coupling occurring for at least two phonon modes of the system. The line width analysis of the phonon modes essentially ruled out the role of magnetostriction effect in the observed phonon anomaly. The investigation of the lattice parameter variation across $T_N$ (obtained from the temperature-dependent neutron diffraction measurements) further confirmed the existence of the spin-phonon coupling.

## Introduction

To meet the increasing need for the next generation spintronic devices, multi-functional materials which respond to various external stimuli, e.g., magnetic field, electric field, pressure, etc. are required. Particular attention has been given to look for the materials showing strongly coupled magnetism and electrical behaviours [1-4]. In this regard, the materials with double perovskite (DP) structure i.e. $R_2BB'O_6$ (R= rare earth and B and B'= transition metal) are of particular importance owing to their diverse fundamental and fascinating physical properties including near room temperature ferromagnetism, large magneto-dielectric effect, giant magneto-resistance, giant exchange bias, magneto-caloric effect etc [3-6]. Hence, such compounds opened up ample opportunities for a broad spectrum of potential applications including sensors, spin filter junctions, memory devices, etc. [7-8]. Depending upon the B/ B' -site ordering, the DPs are found to crystallize mainly into two types of structures: (1) the ordered monoclinic structure with $P_{21}/n$ symmetry (2) the B-site disordered orthorhombic structure with Pnma symmetry [9-10]. Eventually, the ordered DPs show the ferromagnetic (FM) behaviour owing to $180^0$ super-exchange interactions $B^{2+}$-$O^{2-}$-$B'^{4+}$ (typically B= Co/Ni and B′=Mn) but co-existence of $B^{3+}/B'^{3+}$ ions (as a disorder) introduces the competing anti-ferromagnetic (AFM) interactions [5,11]. However, despite the intense research interests and meticulous investigations on such DPs, complete understanding of its electronic structure is still far from well understood. Again, the self-ordering of the B/B' site ions in ordered DP occurs when these two ions have significant difference in their ionic radii and the charge states [9-10]. Conversely, B-site disorder occurs for same charge states of B-site ions. Eventually, site disorder of a system is known to have profound effects on its physical properties which called for rigorous investigations world-wide [5,11-16]. Moreover, few members of the ordered FM DPs exhibit spin-phonon coupling [6,9,16-17]. In fact, coupling between the magnetism (spin) and lattice (phonon) is one of the crucial mechanisms for the coupling between magnetic and electric orders in solids. Thus, to elucidate the magneto-electric coupling phenomena, probing the spin-phonon coupling can be an effective tool. For the strongly correlated oxide systems in which multi-functional properties are observed due to the presence of coupling between different properties, e.g., spin-phonon coupling, electron-phonon coupling etc., Raman spectroscopy can be a sensitive tool for detecting such phenomena [17-19,20].

Eventually, the Co/Ni/Mn-based ordered FM insulating DPs have been extensively studied in the past few years [5,10-11,15]. On the other hand, the studies on the Fe based DPs are relatively limited, and hence much more possibilities are there to explore its physical properties [21-24]. In our previous work in a Fe based DP $Pr_2CoFeO_6$ (PCFO), the B-site disorder was seen to play a crucial role in bringing out a number of interesting phenomena including Griffiths like phase in anti-ferromagnetic background ($T_N$ ~ 269 K) , re-entrant cluster glass behaviour, exchange bias, etc. [25].

However, to date, no comprehensive study is available for spin-phonon coupling in such disordered AFM DPs; thus it can be interesting to study the temperature-dependent Raman spectra to unravel how magnetic ordering affects phonon modes in PCFO. Besides Raman study is also helpful in getting an insight into the B-site cationic ordering in the system.

On the other hand, the local valance states of the B/B′ site ions are strongly correlated to its cationic ordering; thus a prior understanding of its electronic structure can eventually help in further explorations of the origins of its different properties. In the present PCFO system, one end member is $PrCoO_3$ belonging to the cobaltite $RCoO_3$ family which have drawn considerable attention due to its inherent spin degree of freedom in addition to the other degrees of freedom e.g., charge, orbital and lattice [26-28]. This emerges due to the three different energetically (merely) degenerate possible spin configurations of the $Co^{3+}$ ions i.e. low (LS $t_{2g}^6 e_g^0$), intermediate (IS $t_{2g}^5 e_g^1$) and high (HS $t_{2g}^4 e_g^2$) spin states. For $PrCoO_3$, it is still under debate whether the spin state remains in the low spin state (LS) or goes to higher spin states (high or intermediate) up to 300 K [28]. This provoked us to study the electronic structure of PCFO by X-ray photoemission spectroscopy (XPS).

In the present work with PCFO, we have thoroughly investigated the room temperature electronic structure by XPS study and phonon behaviour by temperature dependent Raman measurements.

## Experimental Details

The polycrystalline sample of PCFO was synthesized by the conventional solid-state reaction route which was described elsewhere [25]. The structural analysis yielded a distorted orthorhombic structure (Pnma) [25]. The XPS data was recorded by an Omicron multi-probe surface science system which is well-equipped with a hemispherical electron energy analyzer (EA 125) along with a monochromatic X-ray source Al-$K_\alpha$ line with photon energy 1486.70 eV. The average base pressure was maintained at a value of ~$2.8\times10^{-11}$ Torr. The total energy resolution as calculated from the width of Fermi edge was about 0.25eV. Raman spectra were recorded in a Renishaw inVia Raman spectrometer with 532 nm line of a diode pumped solid state laser delivering power of 5 mW $mm^{-2}$. To control the temperature of the sample, it was kept on a quartz sample holder which was put on the heating/cooling sample cell (THMS-600) connected to temperature control stage (TMS94). The incident laser beam was focused on the sample through the transparent window of the THMS-600 by a 50× short distance objective attached to the Leica DM 2500M microscope. Backscattering geometry was used to collect the scattered beam through the same objective, and a 2400 grooves/mm grating was used as the dispersion element. The spectra were recorded with a spectral resolution of ~1 $cm^{-1}$.

## Results and Discussions

### A. X-ray photoemission spectroscopy study:

The XPS is a widely used versatile and powerful technique to probe the elemental composition and their nominal chemical states of a compound quantitatively. Hence, we have investigated the electronic states near the Fermi level of PCFO by XPS study at 300 K. Fig. 1 (a) is showing the survey spectrum of PCFO system, wherein specifications of the all peaks have been assigned according to the National Institute of Standards and Technology (NIST) database. It reveals that Pr, Co, Fe, and O are present at the surface of PCFO. However, the presence of any other elements except C was not detected from the spectrum, thus confirming the purity of the sample. The presence of C can be attributed to the adventitious molecules absorbed at the surface from the air. We have analyzed the XPS data by correcting the observed binding energies of the elements by referencing the C 1s line at 284.8 eV to eradicate the charging effect. The Fig. 1(b) depicts the core level XPS spectrum of Pr3d region which comprises of two main (spin-orbit coupling) peaks Pr($3d_{5/2}$) and Pr($3d_{3/2}$) at ~933.3 eV and ~953.7 eV respectively. Two additional exchange splitting peaks (ΔE~4.8 eV) originated from the coupling of Pr*4f*, and Pr*3d* hole states are observed at ~928.5 eV and ~948.9 eV [29]. These features are in well agreement with previous reports suggesting a trivalent valance state of Pr [29]. Moreover, for neutral Co atoms, Co 2s peak appears at 925.1 eV which should lie at slightly higher energy for $Co^{3+}$ ions [30]. However, the states Pr $3d_{5/2}$ and Co 2s have a theoretical cross-sectional area ratio of ~8 [30]. Whereas, the integrated areas under the experimental XPS peaks at ~933.3eV and ~928.5 eV have a ratio of ~ 2.59. Thus, if we assume that 933.3 eV peak is arising due to the Pr$3d_{3/2}$ while 927.5 eV peak is due to the Co 2s state, the area ratio of these two curves (~2.59) is in sharp contrast to the theoretically expected ratio (~8). Additionally, the theoretical area ratio of the main Pr3d and its exchange splitting peaks is 2:1 which is close to our observed value (2.59:1). Hence, the observed feature can be ascribed to the predominant contribution from Pr3d states.

Fig. 1(c) depicts the core level Co*2p* XPS spectrum mainly consisting of two spin-orbit splitting peaks centered at ~779.6 eV ($2p_{3/2}$) and ~794.7 eV ($2p_{1/2}$). Two prominent charge transfer satellite peaks are typically observed for the paramagnetic $Co^{2+}$ ions [31-33]. Whereas, mere absence or very weak feature of such satellite peaks in our Co2p XPS spectrum further confirms the presence of octahedrally coordinated $Co^{3+}$ ions [32-33]. Again, the spin-orbit coupling of the observed spectra (ΔE~15.1 eV) is also found to be consistent for $Co^{3+}$ ions (as for $Co^{2+}$ ions, it is ≥16 eV) [33]. Thus, the line shape, peak positions and doublet separation of the observed Co2p XPS spectra indicate trivalent charge state for Co ions, thus supporting our previous results [25].

In Fig. 1(d), the Fe 2p XPS spectra of PCFO have been shown. It comprises two main peaks positioned at ~710.9 eV ($2p_{3/2}$) and ~724.6 eV ($2p_{1/2}$) with ΔE~13.7 eV which are consistent with Fe ions with nominal valance state +3 [34]. Two shake-up satellite peaks marked as $S_1$ and $S_2$ are observed at positions ~718.8 eV and ~732.9 eV respectively. In principle, these satellite peaks appear due to charge transfer between the ligand site ($O^{2-}$) and metal ion $Fe^{x+}$ site. Since, different Fe ions viz., $Fe^{2+}$ and $Fe^{3+}$ have different electronic configurations ($d^6$ and $d^5$), it raises the satellite peaks at different positions (thus treated as hallmark). The Fe2p XPS spectra containing $Fe^{2+}$ ions always show satellite peaks at ~ 6 eV higher than the main peaks whereas that containing $Fe^{3+}$ ions show satellites at ~ 8 eV higher than main peaks [34]. Hence, trivalent Fe ions in PCFO can be inferred.

The core level O 1s spectra have been shown in Fig. 1(e). The spectra consist of a strong peak at ~528.5 eV and a weak feature at ~ 530.8 eV. The first sharp peak at ~528.5 eV is the characteristic feature of the lattice oxygen "$O^{2-}$". However, the small peak at ~530.8 eV is ascribed to the oxygen species that contain fewer electrons due to the adsorption of oxygen leading to the formation of reduced electron rich species $O_2^{2-}$ , $O_2^-$ or $O^-$ [35].

Even though the XPS analysis is a surface sensitive technique but due to having the large mean free path of photoelectrons coming from cobalt (~ 1.5 nm for $E_K^{in}$ =700 eV), the Co*2p* XPS spectra has the ability to provide multiple information viz., charge and spin states, etc. The XPS spectra near the valence band at 300 K have been shown in Fig. 1(f). In the VB spectra containing HS $Co^{3+}$ ions, two distinct peaks are reported to be observed: one at a lower binding energy (near 1-2 eV) and other is at ~ 8 eV [32]. However, for LS $Co^{3+}$, it is theoretically predicted that the spectra should show an intense peak at lower binding energy which is followed by a broad feature upto 8 eV, this is similar to our observed spectra [32]. Hence, the presence of LS $Co^{3+}$ ions can be inferred for PCFO. As a matter of fact, the valence band is mainly composed of O*2p*, Co*3d* and Fe*3d* orbitals [36]. Where, the 3d$t_{2g}$ band lies below the 3d$e_g$ band. Thus, the intense peak $VB_2$ can be mainly ascribed to $t_{2g}$ states of the low spin Co 3d $t_{2g}^6e_g^0$ mixed with high spin Fe 3d $t_{2g}^3e_g^2$ states. On the other hand, the $VB_1$ shoulder is seemingly associated with the $e_g$ states from high spin $Fe^{3+}$ ions (Fe 3d $e_g^2$) which is relatively close to the valence band. The broad feature $VB_3$ observed below $VB_2$ peak is associated to the hybridization of O*2p* and Co/Fe*3d* states and other contributions, e.g. O*2p*-Co/Fe 4*sp* and O*2p*-Pr *5sd* oxygen bonding states. Moreover, the spectral weight near the Fermi level is very weak indicating the Co/Fe*3d* electrons mostly to be localized. This, in turn, predicts the insulating nature of the sample. Thus, the above results are agreeing well with our previous studies [25].

## B. Temperature-dependent Raman study:

The diffraction techniques are helpful for probing the basic crystal structure and its symmetry while the Raman spectroscopy is a unique tool to examine the changes in the local or/and dynamic structural symmetry, local disorder, cationic ordering, etc. [9,10,17-19]. However, for the simple cubic Pm$\overline{3}$m structure of ideal perovskites, all the atoms are centro-symmetrically positioned, and hence no Raman-active phonon modes are found in their Raman spectra. In contrast for the Pnma structure, the distortion caused by the motion of the oxygen atoms around the B-site ions (in $BO_6$) or shifts in R lifts the degeneracy of the Raman modes, and few additional Γ-point phonons become Raman allowed [37]. The symmetry analysis by group theory predicts that there are 60 possible Γ-point phonon modes in such compounds with orthorhombic structure (Pnma). Of these possible modes, only 24 are found to be Raman allowed while among the rest, 25 are infrared allowed, 3 are acoustic translational and 8 are inactive silent modes. The Raman active modes are written in irreducible representation as $\Gamma_g= (7A_g+5B_{1g}+7B_{2g}+5B_{3g})$, where $A_g$ modes correspond to the stretching vibrations of $BO_6$ octahedra and $B_g$ modes refer to the anti-stretching modes of vibration from the same source.

In the Fig. 2 (a), we have shown the experimental Raman spectra (i.e., the scattering intensity variation with the Raman shift) recorded at different temperatures ranging from 300 K to 80 K for $Pr_2CoFeO_6$. It can be seen from the Fig. 2 (a) that of many weak Raman bands, two prominent bands positioned at $\omega_1$~ 435 $cm^{-1}$ and $\omega_2$~ 650 $cm^{-1}$ are observed for all the temperatures. Typically, the phonon modes arising due to the motion of the heavy rare earth ions appear below the 200 $cm^{-1}$, and the modes observed above 300 $cm^{-1}$ are solely due to the motion of light oxygen ions. As already stated the Co/Fe (B-site) ions are centrosymmetric hence they do not contribute to these modes. Thus, both of the observed peaks are associated to the oxygen motion in Co/$FeO_6$ octahedra. Observing the similarities between the observed Raman spectra of PCFO to that of the other earlier reported spectra of perovskites and double perovskites oxides, bands observed at ~435 $cm^{-1}$ and ~650 $cm^{-1}$ can be attributed to the anti-stretching and stretching (or breathing) vibrations of the Co/$FeO_6$ octahedra respectively [9,10,16,19,37-42]. Illiev et al. have explicitly shown by the lattice dynamical calculations that the lower energy band $\omega_1$ involves both the bending and anti-stretching vibrations whereas the higher energy band $\omega_2$ arises purely due to the stretching vibrations [40-41]. It is a comprehensible fact that for the fully disordered DPs; the number of Raman excitation peaks should remain the same as for the single perovskite, i.e. $RBO_3$ or $RB'O_3$. However, the only noticeable change that can be observed is the change in the phonon frequency, and their peak widths arising due to the changes occurred in the average force constants (due to random B-O/ B'-O bonds) and their phonon lifetimes [38,42]. As PCFO has B-site disordered orthorhombic structure, the observed broadness in both of the bands $\omega_1$ and $\omega_2$ can be ascribed to the random site distribution of the Co/Fe ions. Moreover, another feature that can be noted in the peaks (Fig. 2 a) is the asymmetry in its shapes. Eventually, due to the presence of cationic disorder in PCFO, the different B/B′-O vibrations (these unresolved contributions originating from different B/B′-$O_6$ octahedral co-ordinations) will lie close in energy leading to the formation of an asymmetric band envelope.

As a matter of fact, ordered DP systems ($P_{21}/n$) e.g., $La_2CoMnO_6$, $Pr_2CoMnO_6$, $Nd_2CoMnO_6$, $Y_2CoMnO_6$, and $Y_2NiMnO_6$ etc. exhibit additional number of low intensity Raman modes (along with the two main intense stretching and anti-stretching modes) which are attributed to the B-site cationic ordering in the system [6,43,44]. Factually, in the ordered DPs, the effective lattice parameter increases due to the cationic ordering leading to a Brillouin zone folding which essentially gives rise to additional new Γ-point Raman modes [38]. Moreover, for the epitaxial thin film of the DP compounds, a clear peak splitting (which is typically considered as a hallmark for cationic B-site ordering) is observed in both of the stretching and anti-stretching modes [9,10,38]. On the contrary, for the compounds with disordered orthorhombic (Pnma) structures show less number of Raman modes as compared to that for ordered monoclinic ($P_{21}/n$) structures [19, 39,43]. As a matter of fact, we could identify only two main broad peaks $\omega_1$~435 $cm^{-1}$ and $\omega_2$~650 $cm^{-1}$ for PCFO, suggesting towards the random site distribution of the Co/Fe atoms which well accord to our previous work [25].

On the other hand, the DPs containing R-site ions of large ionic radii usually show smaller number of Raman modes. This is because large R-ions cause smaller octahedral tilts and relatively small distortions which lead to the rise of Raman modes at low wave numbers but of weaker intensities as compared to that of DPs with smaller R-ions (causing large octahedral distortions). Thus, for DPs with smaller R-ionic radii show more resolvable Raman modes as compared to the DPs with large ionic radii [43,44]. Hence, the absence of such additional Raman modes for PCFO can also be elucidated based on the large ionic radii of the Pr-ions.

Raman spectra at different temperatures ranging from 300 K down to 80 K have been recorded to study the effect of magnetic ordering upon the lattice vibration (Fig. 1a). To explore the impact of the temperature on the relative intensities of the two Raman modes (stretching and anti-stretching), we have normalized each of the Raman spectra concerning the most intense peak. Interestingly, the intensities of the two bands $\omega_1$ and $\omega_2$ vary differently with temperature. The intensity of the most intense peak at room temperature, i.e., the anti-stretching mode ($\omega_1$) remains almost constant throughout the temperature range whereas that of stretching mode ($\omega_2$) keeps increasing with decreasing temperature. This behaviour is scarce and interesting since the Raman intensity usually increases with decreasing temperatures owing to the reduced phonon scattering at low temperatures. Fig. 2(b) shows the plot of relative intensity $|I_{\omega 1}- I_{\omega 2}|$ as a function of temperature. Interestingly, as the temperature reaches $T_N$ ~269 K, the intensities of both the Raman modes $\omega_1$ and $\omega_2$ become merely equal. Below 269 K, the intensity of $\omega_2$ peak starts dominating over that of the $\omega_1$ peak. This indicates that the onset of magnetic ordering is affecting the phonon modes, thus leading to this anomalous behaviour of their thermal variation of intensities.

To further confirm whether the magnetic spin ordering affects the lattice vibrations, we have investigated the "temperature variation of the Raman excitation frequencies for both the modes i.e. $\omega_1$ and $\omega_2$" (Fig. 2 c & d). Since, for the whole temperature range, no additional Raman mode could be observed, it undoubtedly ruled out the possibility of any global structural transition. Under such condition, the temperature variation of the phonon excitation wave number should follow the anharmonic behaviour (due to usual thermal lattice contraction) which is described by the following expression [45]:

$$\omega_{anh}\,(T) = \omega_0 - C\left(1 + \frac{2}{e^{\frac{\hbar\omega_0}{K_B T}} - 1}\right);$$

Where $\omega_0$ and $C$ are the adjustable parameters, T is the temperature, ħ is reduced Planck's constant and $K_B$ is the Boltzmann's constant. According to this function, the phonon frequency of a particular mode should exhibit gradual hardening with decreasing temperature and reach a plateau at sufficiently low temperatures. From Fig. 2(c-d), such phonon mode hardening down to the temperature $T_N$ ~ 270 K can be observed for both of the stretching and anti-stretching modes. Thus, it clearly suggests that anharmonicity is playing a dominant role in the temperature dependence of PCFO's phonon modes. Interestingly, it is discernible from Fig. 2(c-d) that both the stretching and anti-stretching modes, i.e., $\omega_2$ and $\omega_1$ deviate from the anharmonic behaviour below magnetic ordering temperature $T_N$. However, for the stretching mode $\omega_2$, the theoretical curve attains the plateau region after going down to sufficiently low temperature < 150 K above which it continues the hardening while the experimental curve shows a dramatic slope change and exhibits anomalous softening below $T_N$ (Fig. 2c). On the other hand, as evident from Fig. 2(d) the anti-stretching mode is also largely deviating from anharmonic behaviour by showing clear anomalous softening below $T_N$ which is typically observed due to the spin-phonon coupling in other systems [6,16,41-44, 46-48]. Thus, these observations of anomalous softening of phonon modes (near $T_N$) involving vibrations of magnetic Fe ions indicate towards effective modulation in the lattice vibrations due to the magnetic ordering, thus leading to the spin-phonon coupling. However, in most of the perovskites compounds exhibiting SP effect, such anomalous phonon softening was observed only in the stretching mode. On the contrary, the above results suggest that for PCFO at least two phonon modes are exhibiting the spin-phonon coupling and thus it eventually places this system amongst the rare materials.

It is pertinent to mention here that magnetostriction effect can also affect the phonon frequency (by altering the unit cell volume and lattice constants) leading to the anomaly in its temperature variation [39, 43,44,47-49]. Thus, it is crucial to confirm whether it is the spin-phonon coupling or the magnetostriction which one is causing the observed phonon anomaly. As a matter of fact, the full width at half maximum (FWHM) or line width of the relevant phonon modes remain unaffected by the subtle changes occurred in the volume/lattice parameters caused by magnetostriction effect. On the contrary, FWHM is related to the process of the phonon delay (i.e., lifetime) and thus can be affected by the spin-phonon coupling showing anomaly across magnetic transition [39,48]. Hence, we have investigated the thermal variation of the FWHM of the two modes ($\omega_2$ and $\omega_1$) as shown in Fig. 3 (a-b). Interestingly, for both of the modes, their FWHM showed an anomalous increase below $T_N$~270 K and thus showed clear deviation from the expected anharmonic behaviour (i.e., monotonous decrease in phonon lifetime/FWHM with decreasing temperature due to anharmonic perturbations) [39,48,49]. In this case, the observed anomaly can be realized by the increase in the phonon lifetime beyond $T_N$ which is related to the decrease in phonon energy. In general, the phonon lifetimes are changed mainly due to two reasons (1) spin-phonon coupling and (2) electron-phonon coupling [20,39,48,49]. However, the present XPS study and previous other experimental and theoretical studies performed on this system showed it to be an insulator which readily discards the possibility of electron-phonon coupling [25]. Thus, the observation of FWHM anomaly can univocally be ascribed to the spin-phonon coupling in PCFO.

However, to further ascertain the absence of the role of magnetostriction effect in the observed phonon anomaly, we have investigated the temperature variation of the unit cell volume and the lattice parameters (Fig. 3d). The lattice parameters were obtained by refining the temperature dependent the neutron diffraction (ND) data by Rietveld method (Fig. 3c). It can be noted that a magnetic reflection (011) (marked by *) starts appearing at ~$16^0$ below $T_N$ which is suggesting the onset of long-range ordering in PCFO. The magnetostriction effect is typically manifested in the form of a remarkable anomaly in the volume and/or lattice parameter data across $T_N$, which provides a shred of direct evidence for this effect [48-50]. However, as evident from the Fig. 3d, no such remarkable anomaly can be observed in the volume and lattice parameters curves. Hence, it directly rules out the possibility of magnetostriction effect causing the observed phonon anomaly across $T_N$. Therefore, all these above facts unambiguously establish the presence of the strong interplay among the microscopic degrees of freedom viz., spin and phonon in this system.

Moreover, according to the mean field approximation when the long-range magnetic ordering sets in a structure, it induces a renormalization of the phonon frequency [18,19,44,46,47]. The phonon renormalization is found to be proportional to spin-spin correlation function ($< S_i.S_j >$), where $S_i$ and $S_j$ are denoting the nearest neighbour spins situated at ith and jth sites. The change in the phonon frequency due to this renormalization follows the same trend as the normalized magnetization leading the following formula:

$$\delta\omega(T) = \omega(T) - \omega_{anh}(T) = \left(\frac{M^2(T)}{M_0^2}\right)$$

Where M(T) represents the average magnetization at temperature T and $M_0$ is the saturation magnetization. The difference between the observed and theoretical anharmonic fitted data is denoted by $\delta\omega(T)$. Hence, to further investigate the spin-phonon coupling, we have plotted the temperature variation of $\delta\omega(T)$ and $(M(T)/M_0)^2$ for both the phonon modes (Fig. 4 a-b). As evident from the figures, $\delta\omega(T)$ and $(M(T)/M_0)^2$ follow a similar trend showing a downturn nearly at the same temperature ($T_N$~270 K). This is a clear indication that there is a strong interplay between these two behaviours. However, for both the modes, $\delta\omega(T)$ and $(M(T)/M_0)^2$ do not overlap with each other. This can be presumably attributed to the existence of multiple magnetic phases in the system viz., AFM, FM, and spin glass phases which contributes towards different exchange interactions (both with values and signs) which in turn contribute differently in phonon renormalization [25,43,44,47]. Similar deviations from mean field theory were observed in few other systems viz., $Y_2CoMnO_6$, $La_2CoMnO_6$, $Sr_{0.6}Ba_{0.4}MnO_3$, etc. where complex FM/AFM and/or spin glass states co-exist, thus adding more complexity in the spin-phonon coupling process [43,44,47]. However, further study on its single crystal or epitaxial thin film may be helpful to understand more about the intriguing underlying physics in it.

## Conclusions

To summarize, we have investigated the electronic structure of PCFO near the Fermi level by analyzing the XPS spectra at 300 K. The analysis yielded the nominal oxidation states of all the three cations viz., Pr, Co, and Fe to be +3 which in turn predicts towards a B-site disordered structure of PCFO. Valence band spectra analysis showed the absence of electronic states at the Fermi level, thus suggesting an insulating nature of the system which supported our previous report. The Raman spectra analysis yielded a random distribution of B-site ions (Co/Fe) which is primarily triggered by the same charge states of the relevant ions. Moreover, the anomalous behaviour observed in the relative intensities of the stretching and anti-stretching Raman modes across $T_N$ indicated the interplay of the spins and phonons in PCFO. Interestingly, both the Raman modes showed anomalous softening below $T_N$ and deviated from anharmonic behaviour, thus confirming the existence of spin-phonon coupling for at least two modes. Besides, the temperature variation of the line widths of these two modes also showed remarkable anomaly below $T_N$ which gave further confirmation of the spin-phonon coupling. Again, no anomaly in the lattice volume or lattice constants (a, b and c) across $T_N$ could be observed in their temperature variation as obtained from neutron diffraction data analysis. This unambiguously confirmed the absence of magnetostriction effect in the observed phonon anomaly, thus, in turn, established the existence of the spin-phonon coupling in the system. Observation of such spin-phonon coupling for at least two phonon modes is very scarce for such compounds which eventually place it amongst the rare materials.

## Acknowledgments

The authors are grateful to the DST, India for funding the research facilities in the department of physics, IIT (BHU) Varanasi.

Figure Captions:

Figure. 1: (a) depicts the survey scan. (b), (c),(d),(e) and (f) are showing the Pr*3d*, Co*2p*, Fe*2p*, O*1s* and valance band spectra for PCFO.

Figure. 2: (a) shows the Raman spectra at different temperatures. (b) shows the relative height variation of two modes with temperature. (c) and (d) are showing anharmonic fit to the "thermal variation of Raman shift" curves for stretching and anti-stretching modes respectively.

Figure. 3: (a) and (b) are depicting the FWHM variation with temperature for stretching and anti-stretching modes respectively. (c) shows the ND data at different temperatures. (d) and its inset are showing the thermal variation of lattice volume and lattice constants (a, b/√2 and c) respectively.

Figure. 4: (a) and (b) depict the temperature variation of $\delta\omega(T)$ and $(M(T)/M_0)^2$ for the stretching and anti-stretching modes respectively.

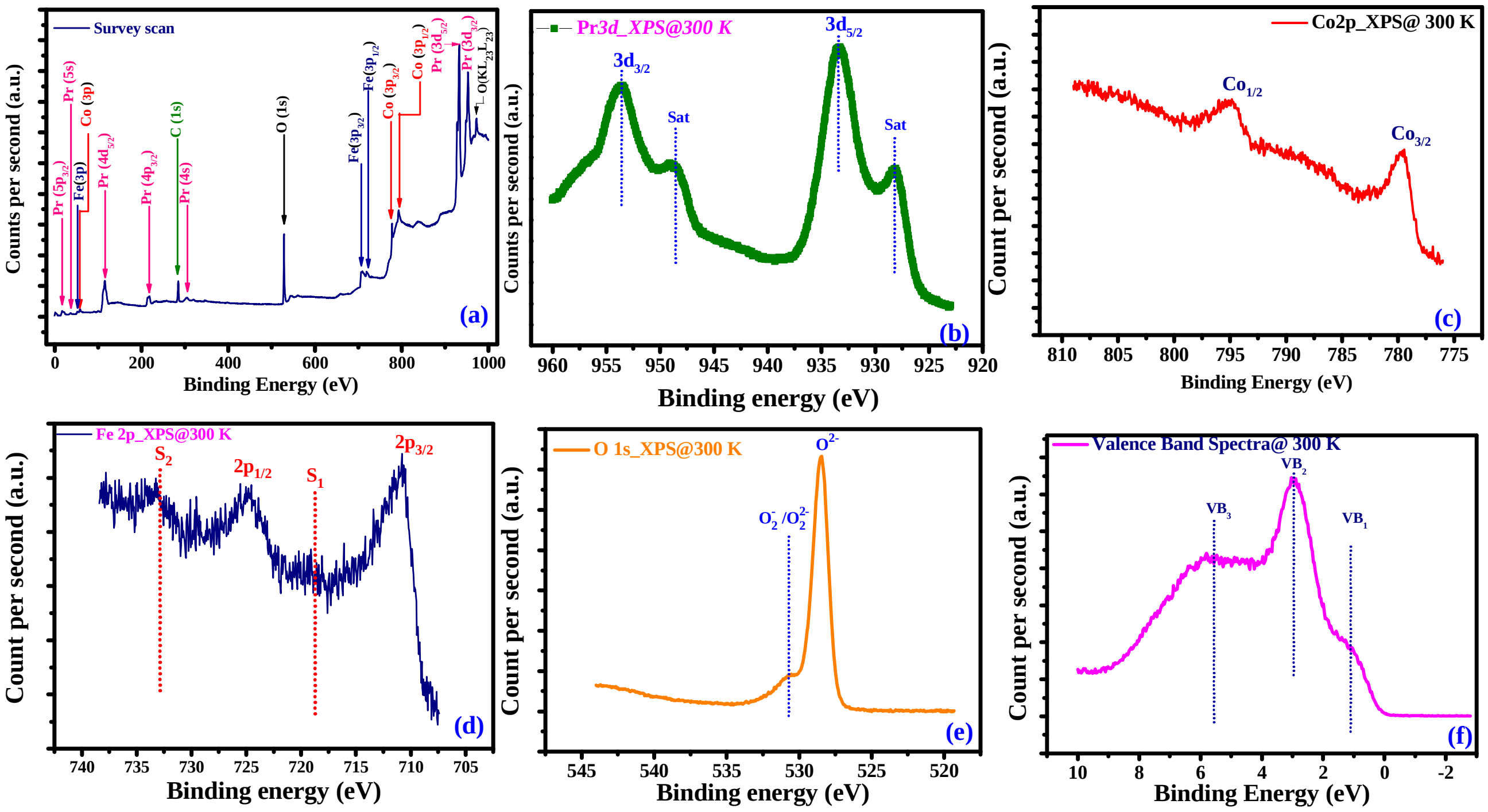

**Fig. 1**

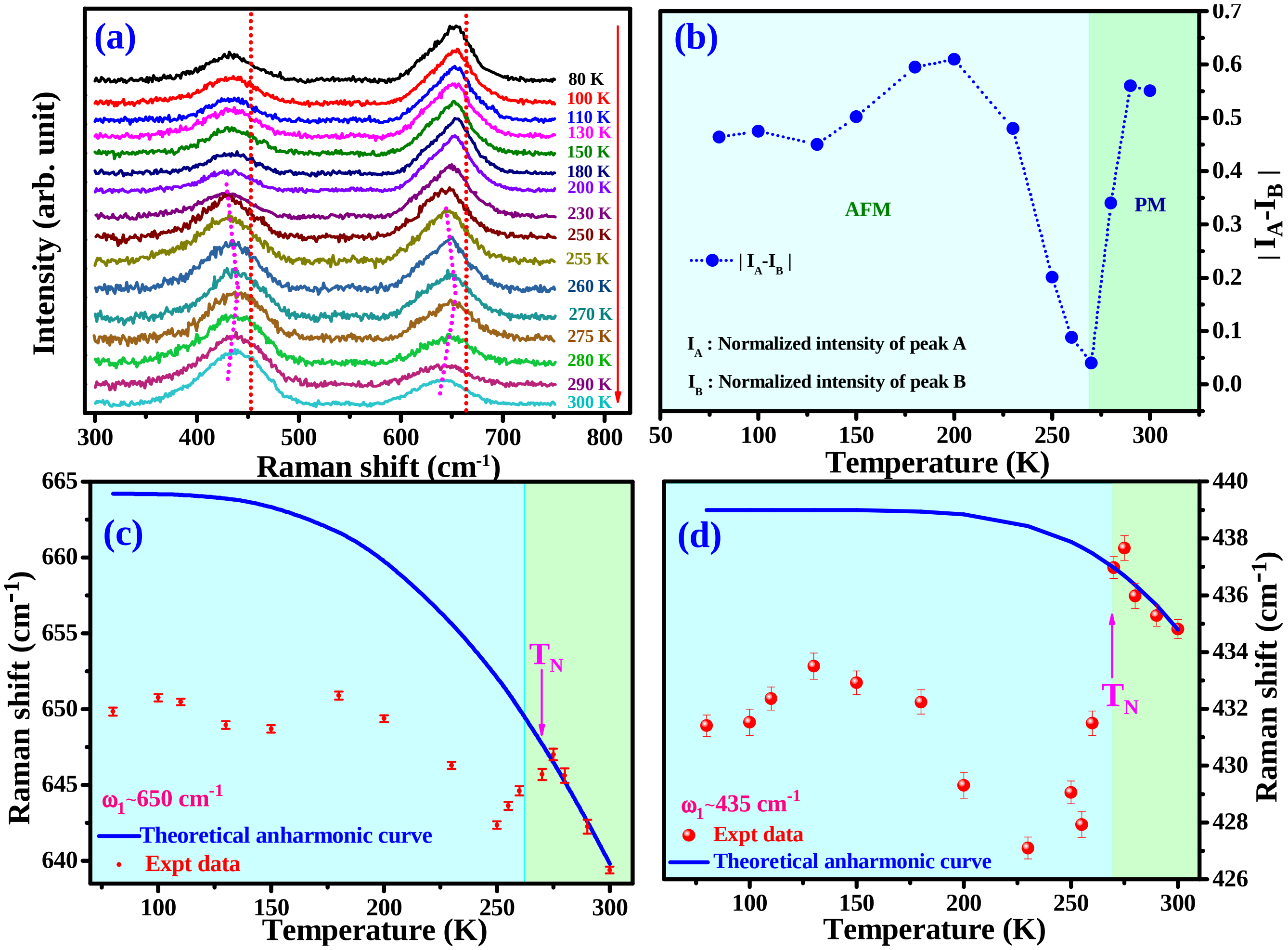
(a)
80 K
100 K
110 K
130 K
150 K
180 K
200 K
230 K
250 K
255 K
260 K
270 K
275 K
280 K
290 K
300 K
Intensity (arb. unit)
Raman shift (cm-1)
(b)
AFM
PM
| IA-IB |
IA : Normalized intensity of peak A
IB : Normalized intensity of peak B
Temperature (K)
(c)
TN
ω1~650 cm-1
Theoretical anharmonic curve
Expt data
Raman shift (cm-1)
Temperature (K)
(d)
TN
ω1~435 cm-1
Expt data
Theoretical anharmonic curve
Raman shift (cm-1)
Temperature (K)

Fig. 2

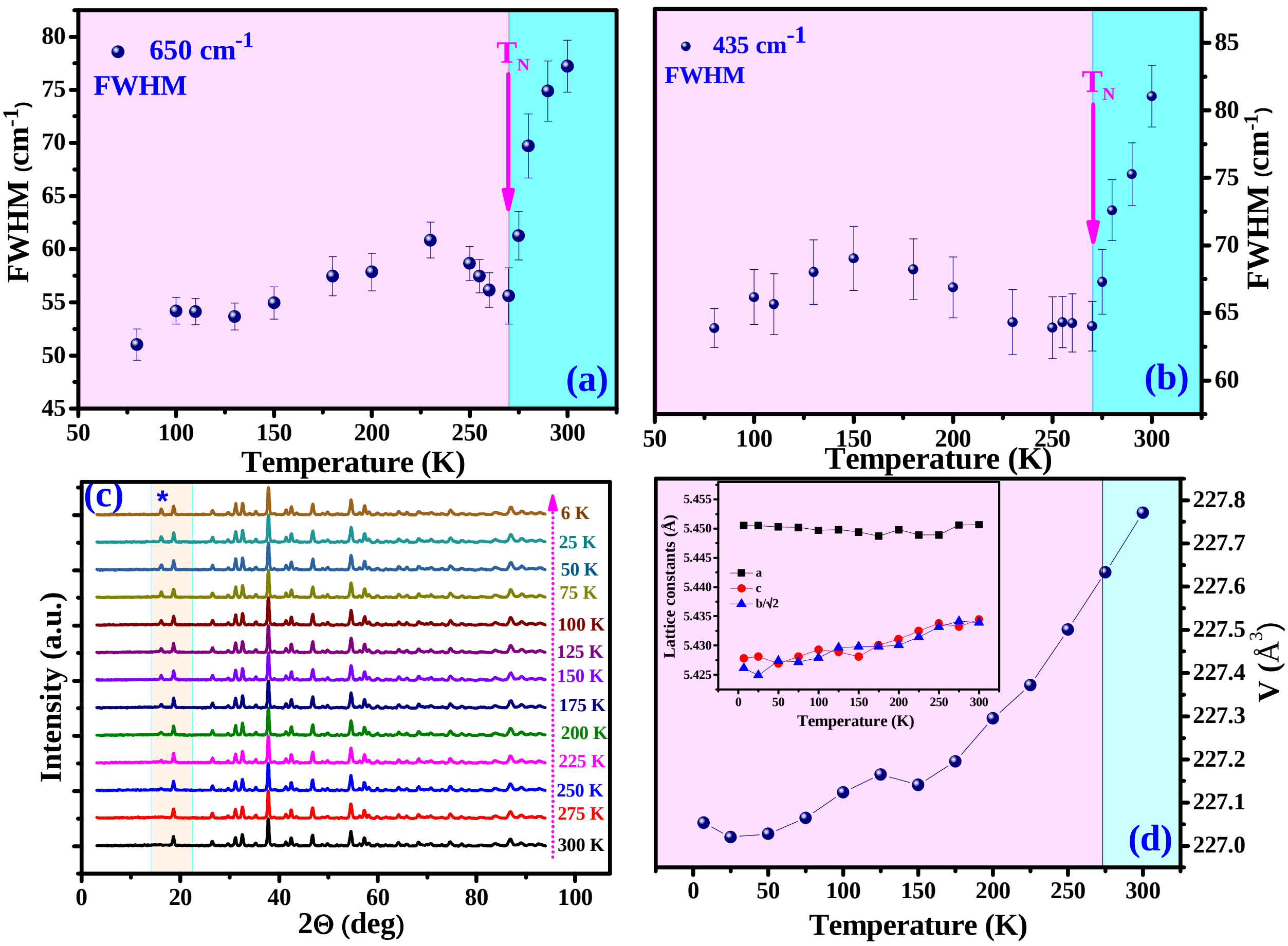

Fig. 3

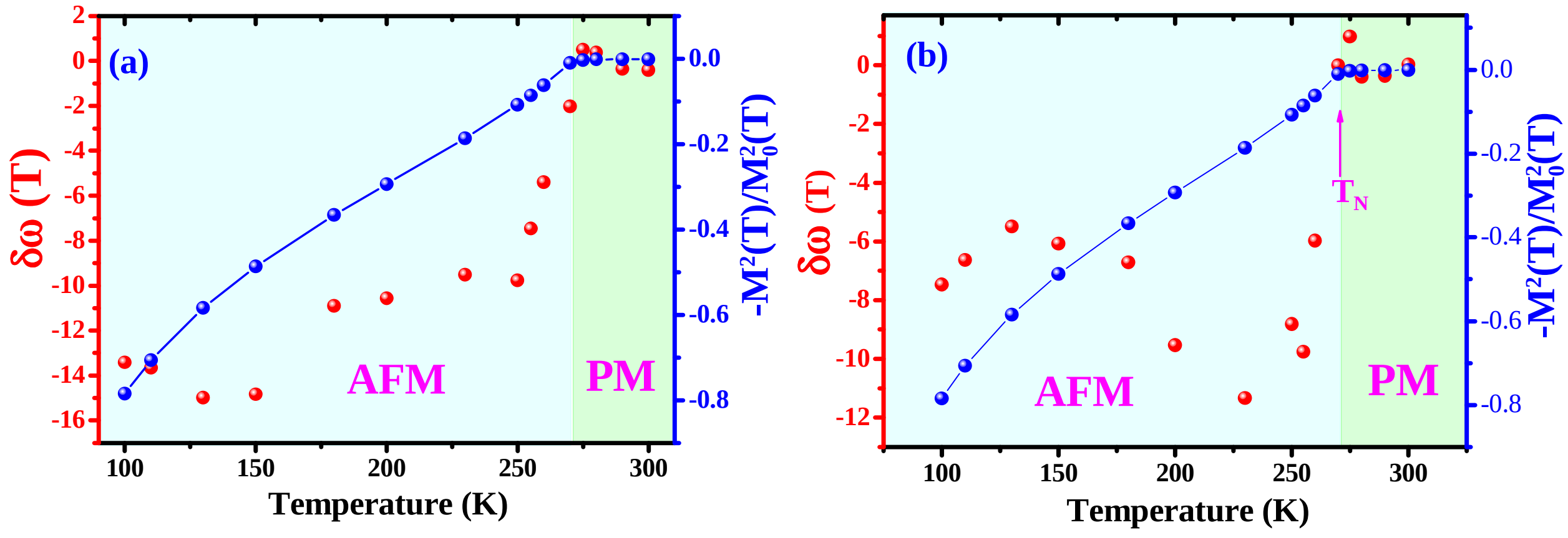

Fig. 4